# Death ligand concentration and the membrane proximal signaling module regulate the type 1/ type 2 choice in apoptotic death signaling

Subhadip Raychaudhuri[1,2,*] and Somkanya C Raychaudhuri[3]

[1] Indraprastha Institute of Information Technology, Delhi, Delhi 110020, India.

[2] Department of Chemistry, University of California Davis, Davis, CA 95776, USA

[3] Department of Biomedical Engineering, University of California Davis, Davis, CA 95776, USA

* Author to whom correspondence should be addressed.

Email: subraychaudhuri@gmail.com

**Abstract**

Apoptotic death pathways are frequently activated by death ligand induction and subsequent activation of the membrane proximal signaling module. Death receptors cluster upon binding to death ligands, leading to formation of a membrane proximal death-inducing-signaling-complex (DISC). In this membrane proximal signalosome, initiator caspases (caspase 8) are processed resulting in activation of both type 1 and type 2 pathways of apoptosis signaling. How the type 1/type 2 choice is made is an important question in the systems biology of apoptosis signaling. In this study, we utilize a Monte Carlo based *in silico* approach to elucidate the role of membrane proximal signaling module in the type 1/type 2 choice of apoptosis signaling. Our results provide crucial mechanistic insights into the formation of DISC signalosome and caspase 8 activation. Increased concentration of death ligands was shown to correlate with increased type 1 activation. We also study the caspase 6 mediated system level feedback activation of apoptosis signaling and its role in the type 1/type 2 choice. Our results clarify the basis of cell-to-cell stochastic variability in apoptosis activation and ramifications of this issue is further discussed in the context of therapies for cancer and neurodegenerative disorders.

**Introduction**

Apoptosis, a programmed mode of cell death, is used in a variety of cellular and physiological situations ranging from developmental regulations to tissue homeostasis. Apoptosis is frequently mediated by death ligands (DL) that are known to cluster death receptors (DR) and membrane proximal adaptor proteins in a signaling complex called DISC (death inducing signaling complex). Procaspase 8 is recruited by adaptor molecules in the DISC and gets cleaved to its active form caspase 8, which in turn, activates both type 1 and type 2 pathways of apoptosis. However, a clear mechanistic understanding of DISC formation and caspase 8 activation, under death ligand induction, is lacking. In addition to its relevance in the biology of apoptosis, such mechanistic understanding can be key to developing therapies for apoptosis related diseases. Death ligand (such as FasL) induced apoptosis has been shown to be the mode of cell death in activation induced cell death (ADCC) in lymphocytes (Elmore 2007). In addition, killing of pathogen infected cells or tumor cells by cytotoxic T lymphocytes (CTLs) and natural killer (NK) cells is frequently mediated by death ligands (Elmore 2007). TRAIL, a death ligand found on the plasma membrane of NK cells (Takeda et al. 2001), has emerged as a promising agent in cancer chemotherapy (Falschlehner et al. 2007; Shirley et al. 2011; Kurita et al. 2011; Picarda et al. 2012). Recent experiments have shown that the fraction of melanoma (skin cancer) stem cells is much higher in NOD SCID IL2rg-/- mice than that found in regular NOD SCID xenotransplantation assays (Quintana et al. 2008). Mice lacking the IL2rg receptor chain are known to be depleted of NK cells, thus, indicating an important role for TRAIL mediated killing of tumor cells. DISC formation and Caspase 8 activation have also been implicated in neurodegenerative disorders such as in Alzheiemer's disease where Aβ (amyloid beta) oligomers and aggregates provide the apoptotic stimuli (Picone et al. 2009; Di Carlo 2010). In addition to apoptosis, death receptor activation has been shown to be involved in other forms of cell death such as necrosis and necroptosis (Kim et al. 2000; Daniels et al. 2005).

Death ligand induction leads to caspase 8 activation and thus could activate both type 1 and type 2 pathways. Understanding the mechanisms of type 1/type 2 regulation remains an important problem in the biology of apoptosis. Some of the initial experiments classified cells as either type 1 or type 2 depending on their mode of activation (Scaffidi et al. 1998; Scaffidi et al. 1999). It was indicated that DISC formation and caspase 8 activation is stronger in type 1 cells (Scaffidi et al. 1998). Even though initial studies found that increased caspase 8 activation can switch the activation from type 2 to type 1 (Scaffidi et al. 1998; Hua et al. 2005), a clear mechanistic understanding of such a correlation was lacking as caspase 8 molecules activate both type 1 and type 2 pathways (Okazaki et al. 2008). Crucial insight into the problem can be provided by considering the kinetic rate constants for reactions involving caspase 8 with its immediate binding partners in type 1 and type 2 pathways. Preferential binding of capsase 8 with Bid (type 2 pathway) explains the type 2 phenotype in cells exhibiting low caspase 8 activation (Raychaudhuri et al. 2008; Raychaudhuri and Raychaudhuri 2013). The levels of initiator caspases (pro-caspase 8 and pro-caspase 9) have also been shown to be key regulators of the type 1/ type 2 choice (Okazaki et al. 2008). In a previous work, we have shown that type 1/type 2 choice is regulated at a systems level (Raychaudhuri and Raychaudhuri 2013). How fast the activation signal propagates through the type 1 and type 2 pathways, to finally activate effector caspases (caspase 3/7), was shown to determine the type 1/type 2 choice. Therefore, it becomes clear that both caspase 8 activation and systems level regulation downstream of caspase 8 activation determine the type 1/type 2 choice. Even though an important regulatory role of active caspase 8 in the type 1/type 2 choice has been established, the mechanism of caspase 8 activation (such as by death ligand induction) remains to be clearly elucidated. In this context, one may ask whether concentration of death ligands can impact the type 1/type 2 choice in apoptosis signaling. One also needs to consider the effect of inherent state of the membrane proximal module on the type 1/type 2 choice. Due to its crucial role in the regulatory mechanisms of apoptosis signaling, the question of type 1/ type 2 choice has important ramifications for cancer and neurodegenerative disorders. We have recently studied the possibility of switching the activation from type 2 to type 1 by targeting the membrane proximal apoptotic signaling module in cancer cells (Raychaudhuri and Raychaudhuri 2013). Recent experiments also explored how a type 2 to type 1 transition can be achieved in the context of TRAIL induced apoptosis (Kurita et al. 2011). The problem of type 1/type 2 choice also has important ramifications for neurodegenerative disorders such as in Alzheimer's disease. Aggregated form of Aβ peptides were shown to activate the type 1 pathway of apoptosis, whereas Aβ oligomers mainly activated the type 2 pathway (Picone et al. 2009). Caspase 8 activation was observed under the action of both Aβ oligomers and aggregates (even though Aβ oligomers may enter the cytosol and mediate its dominant apoptotic effect). Therefore, a mechanistic understanding of the activation of the membrane proximal signaling module and its role in the type 1/type 2 choice, during apoptotic activation, can have profound implications for therapies for cancer and neurodegenerative disorders.

Activation of the membrane proximal apoptotic module, hence the type 1/type 2 choice, is further impacted by a systems level feedback mechanism mediated by caspase 6. It is known that active caspase 3 can proteolytically cleave pro-caspase 6 and generate active caspase 6 molecules. Caspase 6 has effector activities (similar to other effector caspases such as caspase 3/7) but they are also capable of activating caspase 8. Thus caspase 8 activation, upon death ligand induction, can propagate the activation signal through the type 1/type 2 loop network finally leading to activation of caspase 3 -> caspase 6 -> caspase 8 in a systems level feedback loop (Figure 1). Here, we study the effect of caspase 6 mediated feedback activation and elucidate the role of caspase 6 on the type 1/type 2 choice. Caspase 6 has attracted much recent attention in the context of neural cell apoptosis neurodegenerative disorders (Leblanc 2013). However, activation of caspase 6 in some of these cases might be through non-canonical pathways (different from the usual caspase 8 -> caspase 3 -> caspase 6 activation).

Cell-to-cell variability in apoptosis signaling has been established as an important regulatory mechanism in apoptotic cell death (Raychaudhuri et al. 2008; Raychaudhuri 2010; Albeck et al. 2008; Dussmann et al. 2010; Lee et al. 2010). Expression levels of pro- and anti-apoptotic proteins are usually cell-type specific (Lopez-Araiza et al. 2006), but significant cell-to-cell variability (within a population of a given cell type) in protein levels has been observed (Spencer et al. 2009). Such cellular variability in apoptotic molecules can generate single cell variability in apoptotic activation (Spencer et al. 2009; Skommer et al. 2010). However, even when all the cellular parameters are identical, cell-to-cell variability can result from inherently stochastic signaling reactions (Raychaudhuri et al. 2008; Raychaudhuri 2010; Skommer et al. 2010). Interestingly, cell-to-cell inherent variability can have important implications for diseases in which apoptosis is dysregulated (Raychaudhuri et al. 2010; Skommer et al. 2010; Brittain et al. 2010; Skommer et al. 2011a; Skommer et al. 2011b). In a recent work, we have shown that low probability activation of Bax by BH3 only activators (such as Bid) is a mechanism for generating cell-to-cell variability (Raychaudhuri and Das 2013). We have also found that cell-to-cell variability in this situation is remarkably sensitive to the Bcl-2 to Bax ratio (Bcl2 to Bax ratio > 1 was linked to slow apoptotic activation with cell-to-cell stochastic variability). In cancer cells, high Bcl2 to Bax ratio is frequently observed and is a mechanism for apoptosis resistance of cancer cells (Certo et al. 2006). In contrast, in Alzheimer's diseases a recent study has shown that the Bcl2 to Bax ratio (to < 1) can be altered by A$\beta$ peptides to favor apoptotic activation, which can then be reversed by the application of brain-derived neurotrophic factors (BDNF) to protect neural cells (Sun et al. 2012). Therefore, novel chemotherapeutic strategies (in cancer and neurodegenerative disorders) might be possible to design that would rely on modulating cell-to-cell stochastic variability in apoptotic activation. Our previous works have indicated that the problem of type 1/type 2 choice is linked to the issue of cell-to-cell stochastic variability in apoptosis (as activation through the type 2 pathway is frequently associated with stochastic variability while type 1 activation remains deterministic) (Raychaudhuri et al. 2008; Raychaudhuri and Raychaudhuri 2013). Hence, elucidating the role of death ligands and the membrane proximal signaling module in the type 1/type 2 choice can have major implications for understanding cell-to-cell stochastic variability in apoptosis activation and also for apoptosis related diseases.

In this work, we develop a hybrid Monte Carlo (MC) simulation to study the question of type 1/type 2 choice in apoptosis activation. Specifically, we address how the following parameters impact the type 1/type 2 choice: (1) concentration of death ligands and (2) the inherent state of the membrane proximal module. Our results indicate increased death ligand concentration increases type 1 activation and may lead to a type 2 to type 1 transition for a given cell type. The expression levels of pro- and anti-apoptotic proteins in the membrane proximal signaling module are also shown to be crucial in caspase 8 activation and the type 1/type 2 choice. Our study provides mechanistic insights into the process of DISC clustering and caspase 8 activation (upon induction of death ligands). In addition, we consider the systems level feedback mechanism mediated by caspase 6 activation and elucidates its role on type 1/type 2 choice. We discuss the implications of our results for cell-to-cell stochastic variability in apoptotic activation and show how such stochastic variability can be utilized to design therapies for cancer and neurodegenerative disorders.

**Methods**

*The signaling model for apoptotic cell death*

We carried out a computational study of death ligand induced apoptosis using Monte Carlo (MC) simulations. A simplified network model of apoptotic death signaling (Figure 1) was studied utilizing controlled in silico experiments. Both type 1 and type 2 pathways were simulated.

In our model, casapse 8 (initiator caspase) activation initiates signaling through both type 1 and type 2 pathways. Death ligands (such as FasL / TRAIL) undergo binding / unbinding reactions with death receptors with nanomolar affinity. Probability parameters $P_{on}$ and $P_{off}$ are used to simulate binding / unbinding between death ligands and receptors (those presumably vary depending on the receptor type). Ligand binding supposedly induces a conformational change in the death receptors leading to their oligomerization (Scott et al. 2009). In our model, two ligand bound death receptors reduce their free energy ($E_{dd}$ is taken to be -2 $K_BT$ unless specified otherwise) when occupying neighboring lattice sites and thus induce clustering of death receptors. Adaptor proteins (such as FADD / TRADD for Fas receptors) could bind to ligand bound death receptors, with moderately high $10^8$ $M^{-1}$ affinity, using an interaction between their death domains (DD). Affinity for this receptor-adaptor interaction was assumed to be similar to that between the adaptor molecule and pro-caspase 8. Adaptor molecules could also bind to free death receptors albeit with a low probability (affinity ~ 1 $M^{-1}$) so that significant apoptotic activation does not occur without ligand induction. Death effector domains (DED) of the adaptor molecules are known to bind with the DED domains of pro-caspase 8 molecules ($k_{on}$: 3.5 × $10^6$ $M^{-1}$ $s^{-1}$ and $k_{off}$: 0.018 $s^{-1}$) (Hua et al. 2005). However, this interaction takes place only when adaptors are already bound to death receptors. Therefore, ligand binding could induce clustering of death receptors and recruitment of adaptor molecules to clustered death receptors. Procaspase 8 molecules are then recruited to the clustered adaptor proteins to generate the assembly of DISC (death-inducing-signaling-complex), ultimately leading to active caspase 8 molecules through autoprocessing (Peter and Krammer 2003). In this work, DISC formation was simulated using a hybrid simulation scheme between kinetic Monte Carlo model for signaling reactions with an explicit free energy based model for the clustering of death receptors (Raychaudhuri 2013).

Active caspase 8 initiates signaling through both type 1 and type 2 pathways. In the type 1 pathway, caspase 8 directly processes procaspase 3 (effector caspases) to generate active caspase 3. In the type 2 pathway, caspase 8 cleaves Bid to an active form (tBid) which translocates to mitochondria to bind with Bax. We have earlier shown that higher affinity of active caspase 8 for Bid, compared with that for pro-caspase 3, leads to preferential activation of the type 2 pathway (at early times) (Raychaudhuri and Raychaudhuri 2013). Additional complexities, in the initial type 1/type 2 choice, might arise from cell type specific rate of Bid cleavage (such as due to Bid phosphorylation status in a given cell type) (Ozoren and El-Deiry 2002), but are not considered here. When two Bax molecules are bound to tBid (on the mitochondrial membrane) they could detach as an active Bax dimer. Apoptotic inhibitor Bcl2 molecules bind with tBid and Bax and thereby inhibit formation of active Bax dimers. It is also possible for Bid to directly activate Bax albeit with a low probability (Raychaudhuri and Das 2013). Cytochrome c is released into the cytosol in an all-or-none manner when the number of active Bax dimers reaches a pre-assigned threshold value (Goldstein et al. 2000; Dussmann et al. 2010). Cytochrome c release leads to cytochorme c-Apaf binding and the subsequent formation of multi-molecular cyto c-Apaf-ATP complex apoptosome. Formation of the apoptosome complex is modeled in a simplified manner (by a cyto c-Apaf-Apaf-cyto c complex) where Apaf represents the Apaf-ATP complex. In our Monte Carlo simulations, effective low probability of apoptosome formation generates stochastic variability in apoptotic activation, but once formed, induces rapid activation of downstream caspases 9 and 3 (Raychaudhuri et al. 2008; Lee et al. 2010; Bagci et al. 2006). Caspase 9 is an initiator caspase (similar to caspase 8), which gets activated in the clustered assembly of apoptosome (whereas caspase 8 gets activated in DISC). The CARD domain of pro-caspase 9 interacts with the Apaf CARD domain (Sheridan et al. 1997; Pan et al. 1997; Mantovani et al. 2001). Active caspase 9, in turn, cleaves procaspase 3 to its active form caspase 3. Apoptotic inhibitor XIAP binds to procaspase 9, caspase 9, and caspase 3 using its BIR domains (Riedl et al. 2001; Shiozaki et al. 2003), thus it could inhibit apoptotic activation in both type 1 and type 2 pathways. However, the release of mitochondrial Smac can antagonize the inhibition of XIAP (Huang et al. 2003; Sun et al. 2002). Smac is released simultaneously with cytochrome c in an all-or-none type manner. Active Caspase 3 can process pro-caspase 6 to proteolytically active caspase 6, another effector caspase but it also has the ability to activate pro-caspase 8 (Sun et al. 2002). Thus caspase 6 provides a systems level feedback loop for both type1 and type 2 apoptotic activation. The details of all the reaction moves that are downstream of caspase 8 and considered in our current simulation (of the apoptosis pathway) are provided in our earlier works (Raychaudhuri and Raychaudhuri 2013). Activation of caspase 3 (effector caspase) closes the type 1/type 2 activation loop (that is initiated by active caspase 8) and thus can be taken to be a downstream readout of apoptotic cell death signaling; MC simulations were carried out to measure the time-course of caspase activation at a single cell level.

Even though the present model can capture some of the essential systems level regulatory mechanisms of the apoptotic signaling pathway, it has several simplifying assumptions. In our model, effects of functionally similar proteins are coarse-grained by a representative protein. Bcl-2 (B cell lymphoma protein 2), for example, represents all the Bcl-2 family proteins (such as Bcl-2, Bcl-xL, Mcl-1) with similar anti-apoptotic properties. In this context, it should be noted that the death receptors / ligands simulated here capture the effect of only one type of receptor / ligand (such as Fas / FasL or DR4 / TRAIL), not the combined effect of all the death receptors / ligands. In our present study, DISC formation has been modeled in a simplified manner. Different isoforms of cFLIP have been implicated in differential regulation of apoptotic activation (Safa and Pollok 2011; Fricker et al. 2010). A role of cleavage product of cFLIP (such as p43-FLIP) in NF-kB activation has also been studied using mathematical modeling (Neumann et al. 2010). Here, we consider only the anti-apoptotic action of cFLIP (such as by cFLIPs that competitively binds with adaptor proteins in the DISC signalosome). In our study, the threshold of apoptotic activation, such as the threshold death ligand needed to trigger apoptosis, is governed by the level of apoptotic inhibitors such as cFLIP, Bcl2 and XIAP. In addition, the rate of synthesis and degradation of signaling molecules (such as caspases) can modify the threshold of activation (Gu et al. 2011) but is not considered here.

*Monte Carlo Simulation of Cell Death Signaling*

Monte Carlo approach has been shown to capture some of the complexities of signaling reactions such as the effect of spatial heterogeneity (Raychaudhuri 2013). Each run of our Monte Carlo simulation corresponds to apoptotic activation in a single cell, thus, Monte Carlo can capture cell-to-cell stochastic variability including inherent variability. We utilize a hybrid Monte Carlo simulation scheme that combines the following approaches: (1) a probabilistic rate constant based (implicit free energy) kinetic Monte Carlo simulation for various reaction moves such as diffusion, binding/unbinding and catalytic cleavage; (2) an explicit free-energy based model that captures clustering of ligand-bound death receptors utilizing energy-function based diffusion moves. Both of the above approaches have been utilized in previous works from this lab (Raychaudhuri 2013). At each Monte Carlo (MC) step molecules are randomly sampled N number of times, where N is the total number of molecules (either free or complexed) present in the system. Therefore, at each Monte Carlo (MC) step, one molecule is sampled (on average) once to allow for either diffusion or a reaction move. All the intracellular reaction moves (considered in our current simulations) and the corresponding kinetic rate constants can be found in our earlier works. Diffusion moves are carried out to one of the randomly chosen neighboring sites (4 for membrane bound molecules and 6 for cytosolic molecules) provided the condition of mutual physical exclusion is satisfied. In the kinetic Monte Carlo part of the simulation, once a diffusion/reaction move is randomly sampled it is accepted only if a randomly generated number in [0,1] is less than the pre-defined probability constant for that particular move (otherwise the move is rejected). The detailed balance condition is satisfied through the ratio $P_{on}/P_{off}$, the probability constants for the binding and unbinding reactions, respectively (Raychaudhuri 2013). In the explicit free energy based part of the simulation, diffusion moves are accepted based on Metropolis criterion ($P_{accept} = \min[1,\exp(-\Delta E/K_B T)]$, where $\Delta E$ is the free energy difference between the new and the previous configuration (Newman and G.T. 1999).

*Estimation of Parameter Values (used in Our Monte Carlo Simulation)*

A simulation volume of $1.2 \times 1.2 \times 1.2$ μm$^3$ (corresponding to a $60 \times 60 \times 60$ lattice with lattice spacing $\Delta x \sim 20$ nm) is chosen in such a manner that the number of molecules (for each molecular species) is equal to the nanomolar concentration. Death receptors are placed on one surface of the simulation lattice ($1.2 \times 1.2$ μm$^2$); death ligands are placed on a surface parallel to the surface on which death receptors are placed. Cytochrome c/Smac is initially contained in a mitochondrial volume of $0.36 \times 0.36 \times 0.36$ μm$^3$ (18 × 18 × 18 lattice). Utilization of a small system size significantly reduces the computational cost of the simulation. Each MC step (ΔT) is chosen to be $10^{-4}$ s based on known mobility of cytosolic molecules. This allows us to take the probability of diffusion (P$_{diff}$) for the fastest diffusing species (cytosolic molecules) to a value 0.5 (an approximate diffusion constant can be estimated D ~ (1/2) × P$_{diff}$ × (Δx)$^2$/(ΔT) ~ (1/2) × 0.5 × (Δx)$^2$/(ΔT) ~ 1 μm$^2$/s). Consistent with lower diffusion constants known for membrane proteins, probability of diffusion for molecules on the plasma membrane (or the mitochondrial membrane) is taken an order of magnitude lower than that is used for cytosolic molecules. It is reasonable to expect that the multi-molecular complex apoptosome will have significantly reduced mobility and its P$_{diff}$ is assumed to be zero. Kinetic reaction rates (such as k$_{on}$/k$_{off}$) and molecular concentrations are obtained from values reported in the literature (mostly from (Hua et al. 2005)) and utilized in our previous work (Raychaudhuri et al. 2008; Raychaudhuri and Das 2013; Raychaudhuri and Raychaudhuri 2013) (unless specified otherwise). Even though Hua et al. considered activation by Fas ligands, the binding affinities for other death receptor-ligands are known to be similar (in the nano-molar range). Probabilistic reaction rate constants used in this study are obtained from kinetic rate constants using a previously described parameter-mapping scheme (Raychaudhuri 2013). P$_{off}$ (or P$_{catalysis}$) simulation parameters are obtained by multiplying k$_{off}$ (or k$_{catalysis}$) values by $10^{-4}$ s (1 MC time-step). Probabilistic parameters for association reactions are determined using the following relation: P$_{on}$ = $10^2$ × k$_{on}$ nM$^{-1}$ s$^{-1}$ (Raychaudhuri 2013). A typical simulation is run for $2 \times 10^8$ MC steps. Controlled Monte Carlo experiments are carried out for specific parameter values (such as molecular concentrations). Each run of the simulation corresponds to activation at a single cell level. We define the following parameters for quantitatively assessing the extent of type 1 and type 2 activations: N$_{type1}$/( N$_{type1}$ + N$_{type2}$) and N$_{type2}$/( N$_{type1}$ + N$_{type2}$), where N$_{type1}$ and N$_{type2}$ represents the number of active caspase 3 molecules generated by caspase 8 and caspase 9, respectively.

## Results

*Expression levels of death receptor and other molecules in the membrane proximal signaling module are key determinants of the type 1/type 2 choice in apoptosis*

Expression levels of death receptors (DRs), death adaptor proteins, apoptotic procaspase 8 and anti-apoptotic cFLIP define the inherent state of the membrane proximal signaling module. Death ligand induction perturbs the membrane proximal module leading to clustering of death receptors and adaptor proteins in the form of DISC and subsequent activation of caspase 8. We first consider the effect of variation in death receptor concentration on apoptotic activation keeping other molecules (including the ones in the membrane module) fixed. Death receptor expression is known to be cell type specific though cell-to-cell variability in death receptor level can be significant (Meng et al.

2011). The cell-type specific variability in death receptor expression has recently been shown to be a regulator of the type 1/type 2 choice in apoptotic death signaling (Meng et al. 2011). Here, we explore the mechanistic basis of the observed correlation between death receptor expression and the type 1/type 2 choice in apoptosis signaling. We studied apoptotic activation as the number of death receptors was varied in the following manner: 2, 10 and 100 molecules. Death ligand concentration was kept constant at 10 molecules in these simulations. In Fig. 2 we show the fraction of type 1 activation which increased with increasing expression of death receptor levels. Consistent with this, the time-to-death (averaged over 60 single cells) decreased with increasing level of death receptors: $T_d = 1.08 \times 10^8$ MC steps for DR = 2, $T_d = 0.73 \times 10^8$ MC steps for DR = 10 and $T_d = 0.72 \times 10^8$ MC steps for DR = 100. Type 1 cells are known to have high death receptor expression (SKW 6.4 has ~225 molecules/$\mu m^2$ and H9 has ~190 molecules/$\mu m^2$) (Meng et al. 2011). In our simulation, type 1 activation was observed for lower concentration of death receptors (than found in typical type 1 cells), but we simulate an effective expression of death receptors and some of the anti-apoptotic factors, such as the expression of decoy receptors (Pan et al. 1997), are not explicitly considered. In addition, lipid mediated interactions and other complexities of DISC formation may modify the type 1/type 2 choice. However, increased type 1 activation with increasing expression of death receptor levels, as observed in our *in silico* studies, should be robust. In our simulations, increased death receptor expression led to more rapid and increased generation of active caspase 8 molecules (Fig. 2) and thereby increased type 1 signaling. Previous studies including that carried out in this lab have indicated a key role of active caspase 8 molecules in the type 1/type 2 choice (Scaffidi et al. 1998; Hua et al. 2005; Raychaudhuri et al. 2008; Raychaudhuri and Raychaudhuri 2013). However, strong caspase 8 activation may not always lead to type 1 activation as, for example, type 2 signaling has been observed in caspase 3 deficient MCF-7 cell line despite significant active caspase 8 generation in those cells (Scaffidi et al. 1998).

Consistent with our earlier findings (Raychaudhuri et al. 2008; Raychaudhuri and Raychaudhuri 2013), type 2 activation was accompanied by large cell-to-cell stochastic variability (Fig 3a). Interestingly, cell-to-cell variation in the caspase 8 activation (within a cell population) provides a mechanism for generating cell-to-cell stochastic variability in apoptosis, especially when the death receptor concentration is low (~ 2 molecules) (Fig 3b). Clearly, cell-to-cell variability in caspase 8 activation contributes to cell-to-cell variability in effector caspase activation. In Fig. 3c, we show that the time to initial caspase 3 (10% of total) activation, at the level of single cells, is correlated with the initiation of caspase 8 activation.

It is reasonable to expect that the inherent state of the membrane proximal signaling module is cell type specific and combined effect of all the molecules in the membrane module impact the type 1/ type 2 choice. In our simulations, we varied the number of molecules in the membrane proximal signaling module in the following manner: (1) FADD = 10, cFLIP = 10 and procaspase 8 = 10, (2) FADD = 100, cFLIP = 100 and procaspase 8 = 100 and (3) FADD = 100, cFLIP = 10 and procaspase 8 = 100. Concentrations of both death ligands and death receptors were kept constant at 10 molecules. In our simulations, FADD represents the adaptor proteins that bind to both death receptor and intracellular signaling molecules such as pro-caspase 8 (see Methods).

In Fig. 4, we show the type 1 fraction of activation as the membrane proximal signaling module is varied. Increased type 1 activation correlated well with increased DISC formation and generation of active caspase 8 molecules (Scaffidi et al. 1998). The time-to-death decreased with increasing type 1 activation: $T_d = 4.3 \times 10^7$ MC steps for FADD = 10, cFLIP = 10, procaspase 8 = 10; $T_d = 3.9 \times 10^7$ MC steps or FADD = 100, cFLIP = 100, procaspase 8 = 100 and $T_d = 1.4 \times 10^7$ MC steps for FADD = 100, cFLIP = 10, procaspase 8 = 100.

In our simulations, increased cFLIP level (or the cFLIP/procaspase 8 ratio) led to reduced level of active caspase 8 generation resulting in decreased apoptosis and increased type 2 signaling. When we simulated Jurkat T cell (leukemia cell line) parameters obtained in (Hua et al. 2005), type 2 signaling dominated apoptotic activation. However, when cFLIP level was inhibited, there was marked increase in type 1 activation. Increased cFLIP level has been observed in certain cancer cells and also been correlated with cancer chemoresistance (Safa and Pollok 2011). However, when cFLIP level was low but FADD and procaspase 8 levels were high, fast and robust type 1 activation was observed in our simulations (Fig 4). Thus it might be possible to induce strong apoptotic activation in cancer cells, by application of death ligands, when cFLIP level is low (or inhibited) but other molecules in the membrane module are robustly expressed.

*Increase in death ligand concentration increases type 1 activation and crucially regulates the type 1/type 2 choice*

Concentration of death ligands presumably varies *in vivo* and the effect of such concentration variation needs to be studied in a systematic manner. We carried out controlled in silico experiments where the concentration of death ligands was varied keeping all other parameters constant. Death ligand concentration was varied in the following manner: 2, 10 and 100 molecules. For each of these death ligand concentrations the level of death receptor was varied in a similar manner: 2, 10 and 100 molecules. We observed that increased level of death ligand induction led to enhanced type 1 activation (Fig 5). When both death ligand and death receptor concentrations were high, rapid deterministic activation of the type 1 pathway has been observed. However, for low level of death ligand induction (~ 1-2 molecules), apoptotic activation was dominated by the type 2 pathway with large cell-to-cell variability (Fig. 6). Therefore, even in cells that are known as type 1 cells (equipped with high level of death receptor expression) it might be possible to alter the signaling phenotype from type 1 to type 2 by decreasing the death ligand concentration (Especially when the interaction between death domains of the death receptor and the adaptor molecule is high affinity with $K_d \sim 10$ nM (Miyazaki and Reed 2001)). In a similar manner, large amount of death ligand induction might increase type 1 activation in type 2 cells. However, in type 2 cells having very low expression of death receptors, such as hepatocytes (Meng et al. 2011), it becomes difficult to activate the type 1 pathway and activation of a second initiator caspase (caspase 9) becomes essential for robust activation of effector caspases. For type 2 cells equipped with very low death receptor expression it might still be possible to activate the type 1 pathway in the following manner: (1) when the type 2 pathway is blocked resulting in a slow type 1 activation, (2) by enhancing expression of pro-apoptotic membrane molecules especially the death receptor expression (Meng et al. 2011), (3) by inhibiting anti-apoptotic proteins such as cFLIP or XIAP (Jost et al. 2009; Raychaudhuri and Raychaudhuri 2013). Other cell-type specific mechanisms for type 2 to type 1 transition are possible, such as, removing sequestration of c-MET bound Fas receptors by application of HGF (hepatocyte growth factors) (Accordi et al. 2007) or enhancing the expression of caspase 3 in caspase 3 deficient cell line MCF-7 (Scaffidi et al. 1998). Therefore, it seems reasonable to expect that both type 1 and type 2 pathways can get activated irrespective of cell types.

It is expected that death ligand concentration and the inherent state of the membrane module would regulate the clustering of DISC and thereby govern caspase 8 activation. The parameter ($E_{dd}$) governing the free-energy reduction of two neighboring death-ligand bound receptors should also be a regulator of DISC generation. Varying this free-energy parameter, resulted in altered clustering of death receptors and DISC generation. The effect of death ligand concentration on the type 1 / type 2 choice is frequently mediated by variation in death receptor clustering and DISC generation. Increased clustering was observed in the case of high death ligand level (100 molecules), which seems to correlate well with activation of caspase 8 (Supplemental Fig. 1). The effect of increased death ligand concentration on receptor clustering was more pronounced when the free energy parameter $E_{dd}$ = -3 $K_BT$ (Supplemental Fig. 1b). As mentioned earlier (*Methods* section), $E_{dd}$ is an effective parameter and may vary depending on the cell type. In addition, it might be possible to enhance $E_{dd}$ (Legembre et al. 2005; Thome et al. 2012) selectively in cancer cells and induce apoptosis by death ligand induction or generating DISC formation by some other mechanisms.

*Caspase 6 provides a system level feedback loop for apoptotic pathways and thereby impacts the type 1/type 2 choice*

Caspase 8 activation initiates signaling through the type 1 and type 2 pathways ultimately resulting in activation of effector caspases (caspase 3/7), thus creating a loop network structure at the systems level. Caspase 6 is another effector caspase that is activated by active caspase 3, but once activated it could also activate caspase 8 providing a mechanism for systems level feedback regulation. It is expected that a significant amount of active caspase 3 will be utilized to carry out their effector functions and only a fraction of it will be available for processing pro-caspase 6. We do not explicitly model the effector activities of active caspase 3, instead, a lower rate effective rate constant is assumed for caspase 3 activation of caspase 6. In our simulations, the rate constants for binding ($P_{on}$: between caspase 3 and pro-caspase 6) and catalytic activation ($P_{cat}$: generation of active caspase 6 from caspase3-pro-caspase 6 complex) were taken to be 0.1 times that was used for caspase 3 activation by caspase 9. Similar rate constants were assumed for binding and catalytic activation for caspase 6 – pro-caspase 8 reaction. It is reasonable to expect that, similar to active caspase3, a significant portion of active caspase 6 will engage its effector substrates (such as lamin A) resulting in effective lower rate constants for its activation of caspase 8.

Caspase 6 activation led to increased generation of active caspase 8 molecules and thus favored the type 1 activation. In Fig 7a, we show the type 1 fraction of activation for two different combinations of death ligand and death receptor concentrations, with and without caspase 6. When both death ligand and death receptor concentrations were moderate (~ 10 molecules or higher), rapid generation of active caspase 8 molecules led to fast deterministic activation of the type 1 pathway and the effect of caspase 6 was to slightly reinforce the fast type 1 activation. For low levels of death ligand concentrations (~ 2 molecules), caspase 6 mediated feedback loop assisted in enhancing type 1 activation. For such low death ligand concentrations, especially when death receptor concentration is also low, apoptotic activation was dominated by the type 2 pathway with large cell-to-cell variability and slow activation of caspase 3; Caspase 6 activation was also slow in accordance with slow activation of Caspase 3. Time-course of caspase 6 activation is provided in Fig. 7b. At the level of single cells, caspase 6 activation was prominent only when robust caspase 3 activation was achieved after apoptosome formation and caspase 9 activation. Thus Caspase 6 activation led to strong caspase 8 activation only after capsase 3 activation reached a significant level (at ~ $5 \times 10^7$ - $10^8$ MC steps) (Fig. 7b and 7c). Consistent with our findings, robust activation of both caspase 8 and caspase 3 has been shown to be delayed (~ hour) in type 2 cells (Scaffidi et al. 1998) (though in those experiments apoptotic activation was probed at the population level). In some of our simulations, such as when death ligand ~ 2 molec. and death receptor ~ 100 molec., dominant mode of activation was altered from type 2 to type 1 due to caspase 6 activation (type 1 fraction: 0.37 ± 0.22 for caspase 6 = 0 and type 1 fraction: 0.52 ± 32 for caspase 6 = 30 molec.). Thus caspase 6 inhibition can be used as a strategy to switch the activation from type 1 to type 2 and to increase cell-to-cell variability. Such a strategy can have important ramifications for the recently explored role of caspase 6 inhibition in treatment of neurodegenerative disorders (Leblanc 2013).

**Discussion**

In this work, we elucidate that the concentration of death ligand and the inherent state of the membrane proximal signaling module affect the type 1/type 2 choice in apoptotic activation. We demonstrate that the inherent state of the membrane proximal module, such as the expression level of death receptors, can make cells of a given type prone to either type 1 or type 2 activation. However, concentration of the death ligand also emerges to be a key parameter in the type 1/type2 choice. Increased concentration of death ligand induction increases type 1 activation and may lead to a type 2 to type 1 transition in a given cell type. Our results provide a mechanistic understanding of caspase 6 activation mediated systems level feedback regulation (through caspase 8 activation) and its impact on the type 1/type 2 choice. Results obtained in this study also indicate an important role of cell-to-cell stochastic variability in caspase 8 activation. Such stochastic variability cannot be captured by ODE based models of caspase 8 activation. Majority of previous studies exploring the type 1/ type 2 choice in apoptosis did not consider the effect of cell-to-cell variability in apoptotic activation.

In this study, formation of DISC and capsase 8 activation was modeled in a simplified manner. Additional complexities might arise from variation in the mode of ligand presentation (such as valency) (Huang et al. 1999), explicit simulation of decoy receptors (Pan et al. 1997) and lipid mediated interactions (George and Wu 2012; Song et al. 2007). Presentation of oligomeric death ligands might lead to more robust DISC formation and induction of type 1 activation in type 2 cells such as hepatocytes (Huang et al. 1999). Taken together with our results, strength of death ligand mediated apoptotic stimuli seems to be a crucial regulator of the type 1/type 2 choice, irrespective of cell types (Raychaudhuri 2010). In this context, it should be noted that membrane bound death ligands might be more effective in inducing receptor clustering in the restricted geometry of the 2-dimensional cell-cell contact. Lipid mediated interactions have been implicated in death ligand induced apoptotic activation even though a complete mechanistic understanding is lacking. Some studies indicate that Fas receptors are localized in sphingolipid rich regions of the plasma membrane in type 1 cells (Sanlioglu et al. 2005). In our current model, clustering of ligand bound death receptors was modeled in an effective manner and is governed by a free energy parameter ($E_{dd}$). Explicit simulation of lipid-mediated interactions would allow us to simulate novel strategies to target the membrane proximal signaling modules in cancer cells (Gajate and Mollinedo 2011; Song et al. 2007; Thome et al. 2012; Xiao et al. 2011; Gulbins and Kolesnick 2003).

We considered the effect of cell-type specific variations in the membrane module on death ligand induced apoptotic activation. However, cell-to-cell variations in the expression levels of signaling molecules (such as the level of death receptors (Meng et al. 2011)) can be significant even within a specific cell type. Such cellular variability may arise from epigenetic regulations and stochastic gene expressions. Based on our results it can be inferred that cell-to-cell variations in the membrane module for a given cell population (of the same type) should also impact the type 1/type 2 choice in apoptotic activation. Therefore, even within a cell population of the same type, one cell equipped with higher level of death receptor and membrane proximal pro-apoptotic molecules may activate the type 1 pathway whereas another cell having lower level of death receptor and other pro-apoptotic membrane molecules might exhibit the type 2 signaling phenotype. Interestingly, such type 1/type 2 variations at the level of single cells can also result from inherent stochastic variability, even when all the cellular parameters (such as the concentrations of membrane molecules) remain identical. Cell-to-cell variability in the time-to-caspase 8 activation (resulting from inherent stochastic fluctuations), for example, can impact the type 1/type 2 choice at the level of single cells. The regulation of type 1/ type 2 choice at the level of single cells is an unexpected feature that emerged from stochastic Monte Carlo modeling.

Cell-to-cell stochastic variability has emerged as a key systems level regulatory mechanism in type 2 apoptosis. Thus the question of type 1/type 2 choice, in apoptotic activation, is frequently linked to the choice between stochastic (large cell-to-cell variability) and deterministic activation. In this work, we elucidated that the inherent state of membrane proximal signaling module and concentration of death ligand affect cell-to-cell variability in apoptotic activation. How cell-to-cell variability can be utilized to target the apoptotic pathway in treatment of cancer and neurodegenerative disorders remains to be explored. In our previous studies, we have explored potential options for inducing stochastic to deterministic transition (or at least minimize stochastic variability), selectively in cancer cells (Skommer et al. 2011a; Raychaudhuri and Das 2013; Raychaudhuri and Raychaudhuri 2013). It was also indicated that cell-to-cell inherent variability should help protect normal cells during cancer chemotherapy. In a similar manner, strategies based on utilization of cell-to-cell variability might be effective in protecting cells from apoptosis in neurodegenerative disorders.

Application of death ligands (such as TRAIL), to induce apoptotic activation in cancer cells, has been established as a chemotherapeutic strategy (Picarda et al. 2012; Shirley et al. 2011; Kurita et al. 2011). However, one needs to consider how to selectively target cancer cells and minimize fractional cell killing. In certain cancer cells, increased expression of death receptors and/or decreased amount of decoy receptor can enhance susceptibility of the cancer cell's membrane module to external apoptotic stimuli (Pan et al. 1997; Sheridan et al. 1997). Such a mechanism presumably explains selective killing of cancer cells observed under TRAIL induction. In addition, vulnerability of the membrane module can be selectively induced (or increased) in some types of cancer cells, such as by increasing the death receptor expression by genetic mechanisms (Ho et al. 2010). However, anti-apoptotic molecules in the membrane proximal signaling module, such as cFLIP, are also frequently over-expressed in cancer cells imparting apoptosis resistance in those cancer cells (Safa and Pollok 2011). In those cases, selective killing of cancer cells can be achieved by simultaneous targeting of a susceptible membrane module and inhibition of cFLIP. In this context, recombinant death ligands or monoclonal antibodies to death receptors (Picarda et al. 2012), having low affinity for death receptors, can be utilized to increase specificity of targeting cancer cells. When we simulate apoptotic activation of susceptible cancer cells having increased death receptor expression (~ 100 molecules) but no cFLIP (also over-expressed Bcl2 family pro- and anti-apoptotic proteins), under the application of small amount of death ligands (~ 2 molecules) with low affinity (~ $10^3$ $M^{-1}$ compared to $10^9$ $M^{-1}$ for typical death ligands), rapid type 1 dominant activation was observed (82% cell death in $10^8$ MC steps). In contrast, apoptotic activation in a normal cell with lower death receptor expression (~10 molecules) but significant cFLIP (~ 30 nM) expression was much lower (7% cell death in $10^8$ MC steps). In addition, the slow activation was dominated by the type 2 pathway with large cell-to-cell stochastic variability. There might exist other mechanisms of perturbing the susceptible membrane module of the apoptotic pathway, such as by combined targeting of death receptors and lipid rafts in a synergistic manner, to induce selective activation in cancer cells. Such selective targeting of the cancer cells can be achieved if only weak type 2 activation is induced in normal cells (where slow activation with large cell-to-cell variability should provide protection).

Caspase 6 activation has been frequently observed in neural cells undergoing degeneration and caspase 6 inhibition is emerging as a promising novel strategy in neurodegenerative disorders such as Alzheimer's disease (Leblanc 2013). In this study of death ligand induced apoptosis, we considered systems level feedback regulation mediated by active caspase 6. Even if the apoptotic trigger is located in the intrinsic (type 2) pathway, caspase 6 mediated activation of caspase 8 can start direct processing of pro-caspase 3 (type 1 activation). Here, we showed how the activation can be switched from type 1 to type 2 under caspase 6 inhibition. Such a strategy, based on systems level regulatory mechanisms and cell-to-cell stochastic variability, could protect a large fraction of cells by minimally perturbing the apoptotic pathway and might turn out to be effective under certain situations. How regulation of cell-to-cell variability can be utilized to protect cells in neurodegenerative disorders remains to be explored. Recent studies indicated the possibility of caspase 6 self-activation in the context of neurodegenerative disorders (Leblanc 2013). However, over-pressed caspase 6 was thought to be required (for such self-processing) which would implicate low probability of activation and a role of stochastic variability (Raychaudhuri et al. 2008; Raychaudhuri and Das 2013). *In silico* studies, such as the Monte Carlo simulations carried out in this work, can be utilized to design optimal strategies for targeting of the apoptotic pathway based on elucidation of its systems level regulatory mechanisms.

# Figures

**Figure 1.** Schematic of the apoptotic death signaling network. Death ligand induction activates apoptosis through two distinct pathways: type 1 (intrinsic) and type 2 (extrinsic). The type 1-type 2 signaling loop is initiated by generation of active caspase 8 and ultimately converges on caspase 3/7 activation. Caspase 6 provides a systems level feedback activation.

**Figure 2.** Type 1 activation fraction (upper panel) and caspase 8 activation (lower panel) for increasing death receptor levels. Type 1 activation is estimated when active caspase 3 concentration reaches 90 nM (90% activation). Death receptor concentration is varied in the following manner: 2, 10 and 100 molecules (death ligand concentration is kept fixed at 10 molecules). Type 1 activation correlates well with caspase 8 activation at $T = 10^8$ MC steps (lower panel). For caspase 8, activation is normalized to its maximal value (~ 30 nM). Data is obtained from 60 single cell experiments (Monte Carlo runs).

**Figure 3.** (a) Time-course of caspase 3 activation for low death receptor concentration. Data is shown for 6 representative (type 2 dominant) cells when DR = 2 molecules. Each color corresponds to apoptosis activation for a single cell (Monte Carlo run). (b) Caspase 3 and caspase 8 activation at a single cell level. Data is shown for 3 representative (type 2 dominant) cells (DR = 2 molecules). Caspase 8 activation precedes caspase 3 activation and contributes cell-to-cell variability in apoptosis activation. (c) Time-to-caspase 8 activation is correlated to time-to-caspase 3 activation at a single cell level. Time-to-caspase 8 activation captures the initiation of caspase 8 activation (caspase 8 = 1); Time-to-caspase 3 activation is for initial (10% of total) caspase 3 activation. Data is shown for 60 single cell experiments (Monte Carlo runs). A straight line fit to the data indicates high degree of correlation between initiation of caspase 8 activation and early caspase 3 activation.

**Figure 4.** Type 1 activation fraction (upper panel) and caspase 8 activation (lower panel) as the membrane proximal signaling module is varied. Simulations were carried out for the following parameters: (1) FADD = 10, cFLIP = 10 and procaspase 8 = 10, (2) FADD = 100, cFLIP = 100 and procaspase 8 = 100 and (3) FADD = 100, cFLIP = 10 and procaspase 8 = 100 (DL = 10 molecules and DR = 10 molecules). Type 1 activation is estimated when active caspase 3 concentration reaches 90 nM (90% activation). Type 1 activation correlates well with caspase 8 activation at $T = 5 \times 10^7$ MC steps. For caspase 8, activation is normalized to a value of 30 nM. Data is obtained from 60 single cell experiments (Monte Carlo runs).

**Figure 5.** Type 1 activation fraction for various death ligand concentrations. Death ligand concentration is varied in the following manner: (a) 2 molecules, (b) 10 molecules and (c) 100 molecules. For each of the death ligand concentration death receptor concentration is varied in a similar manner: 2 molecules, 10 molecules and 100 molecules. Type 1 activation is estimated when active caspase 3 concentration reaches 90 nM (90% activation). Data is obtained from 60 single cell experiments (Monte Carlo runs).

**Figure 6.** Time-course of caspase 3 activation for low death ligand concentration. Data is shown for 6 representative (type 2 dominant) cells for DL = 2 molecules (DR = 100 molecules). Each color corresponds to apoptosis activation for a single cell (Monte Carlo run).

**Figure 7.** (a) Effect of caspase 6 activation on type 1 mode of activation. Type 1 activation is estimated when active caspase 3 concentration reaches 90 nM (90% activation). Data is shown for two different receptor-ligand concentrations: (1) DL=2 molecules, DR=2 molecules and (2) DL=10 molecules, DR=10 molecules (with and without caspase 6; for each receptor-ligand concentration histogram on the right is with caspase 6). (b) Activation of caspase 6 as the receptor-ligand concentration is varied. Data is shown for three different time-points (T = $5 \times 10^7$, T = $1 \times 10^8$ and T = $2 \times 10^8$ MC steps) and two different receptor-ligand concentrations: (1) DL=2 molecules, DR=2 molecules and (2) DL=10 molecules, DR=10 molecules. Activation of caspase 6 is normalized to its maximal value (~ 30 nM). Data is obtained from 60 single cell experiments (Monte Carlo runs). (c) Activation of caspase 8 (upper panel) and caspase 3 (lower panel) with and without caspase 6. Data is shown for three different time-points (T = $5 \times 10^7$, T = $1 \times 10^8$ and T = $2 \times 10^8$ MC steps) and two different receptor-ligand concentrations: (1) left panel: DL=2 molecules, DR=2 molecules and (2) right panel: DL=10 molecules, DR=10 molecules. Activation is normalized to respective maximal values (30 nM for caspase 8 and 100 nM for caspase 3). Data is obtained from 60 single cell experiments (Monte Carlo runs).

Figure 1.

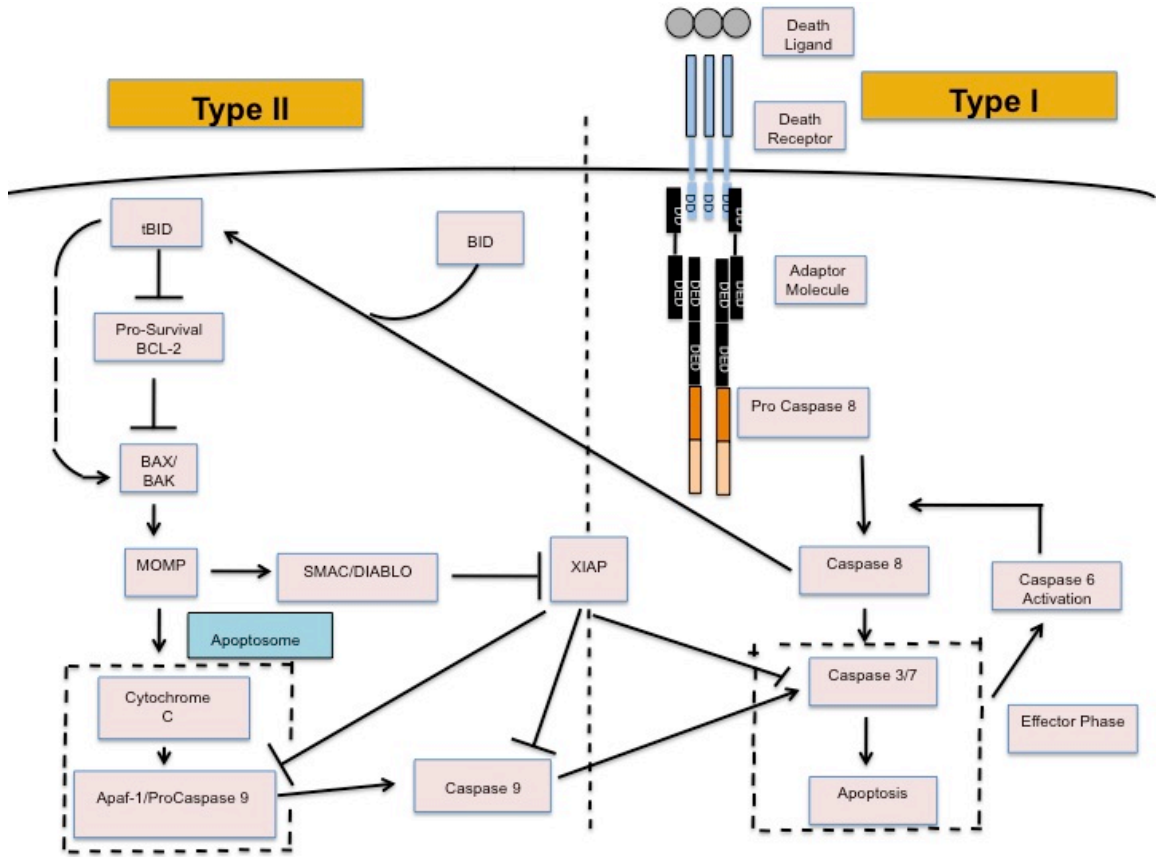

Figure 2.

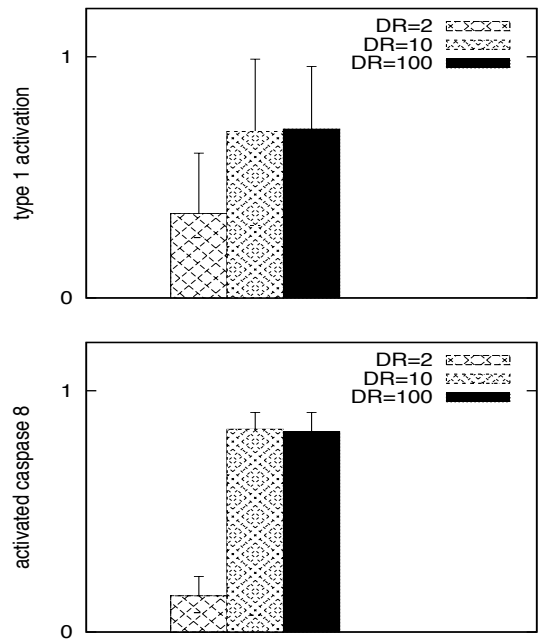

Figure 3a.

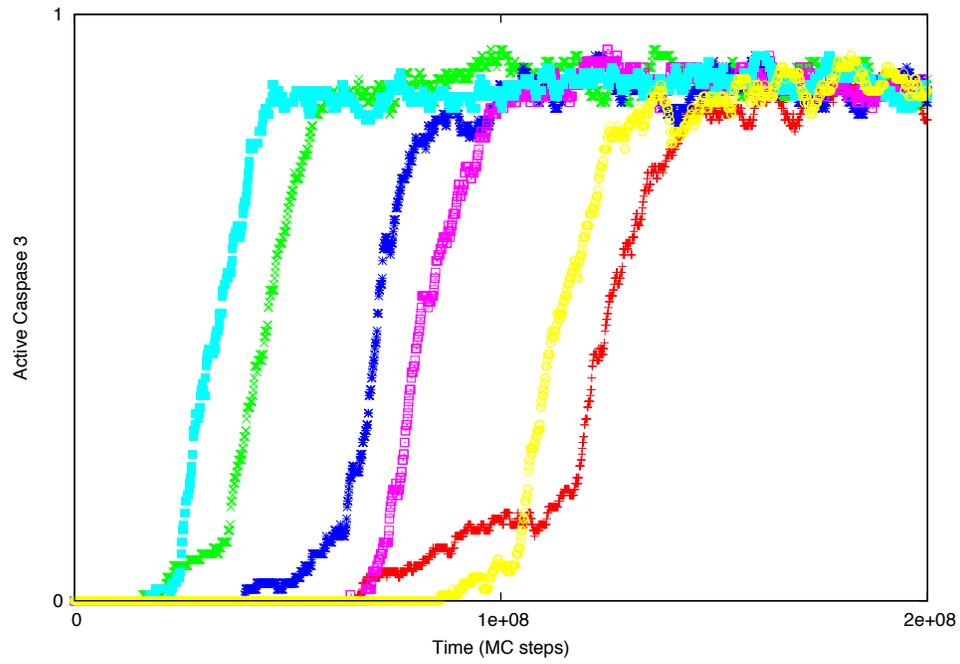

Figure 3b

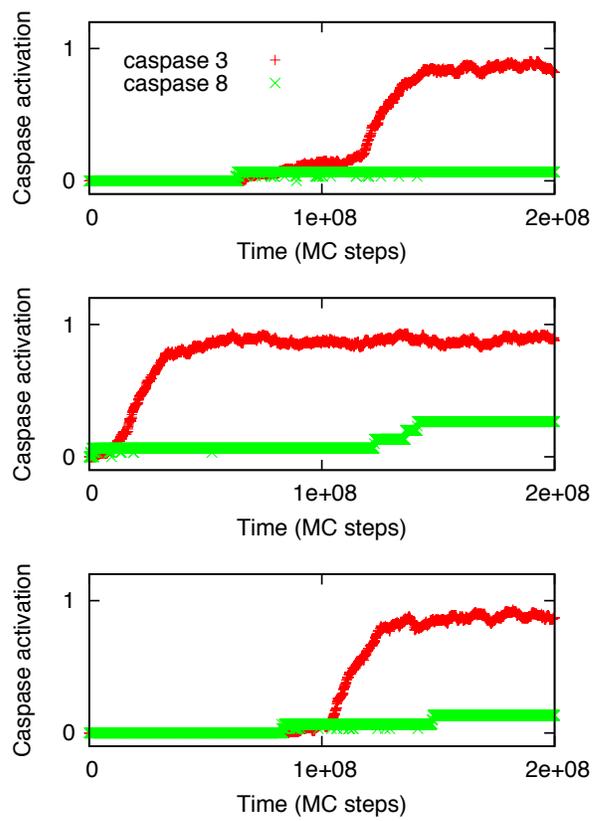

Figure 3c.

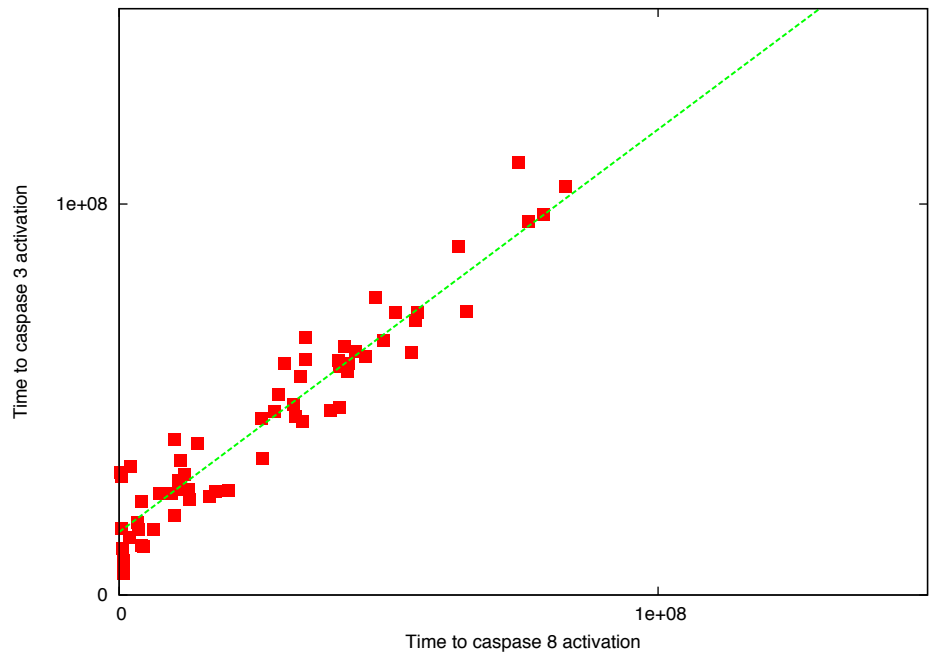

Figure 4.

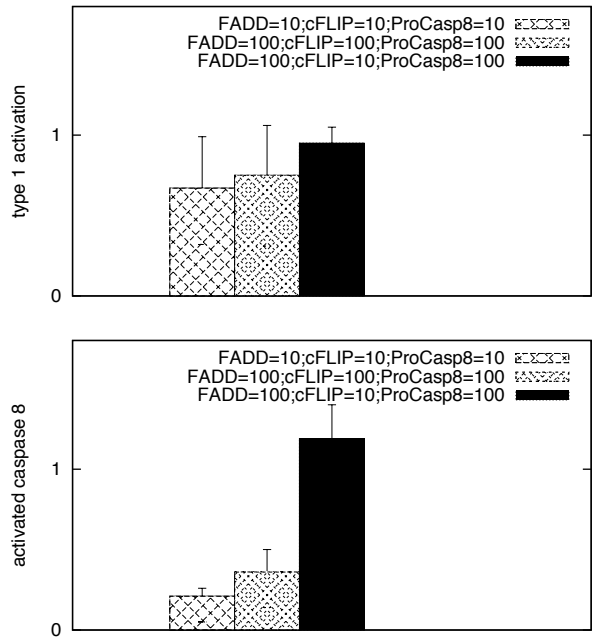

Figure 5.

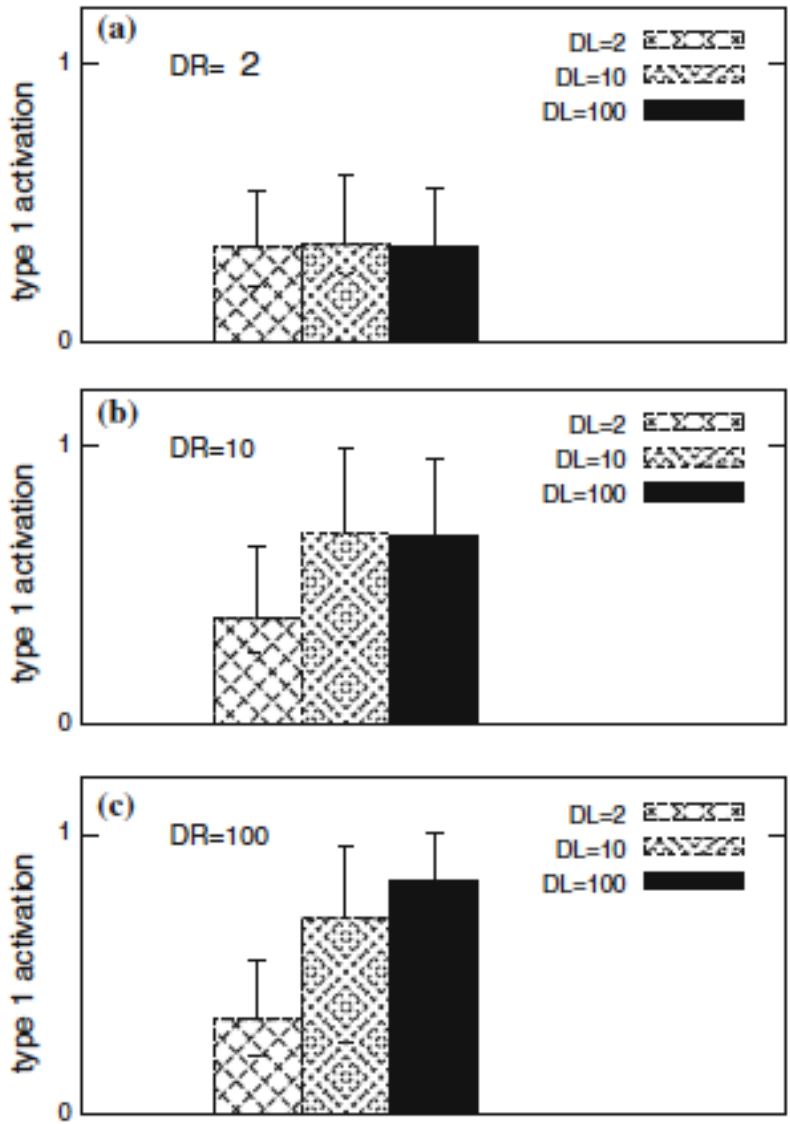

Figure 6.

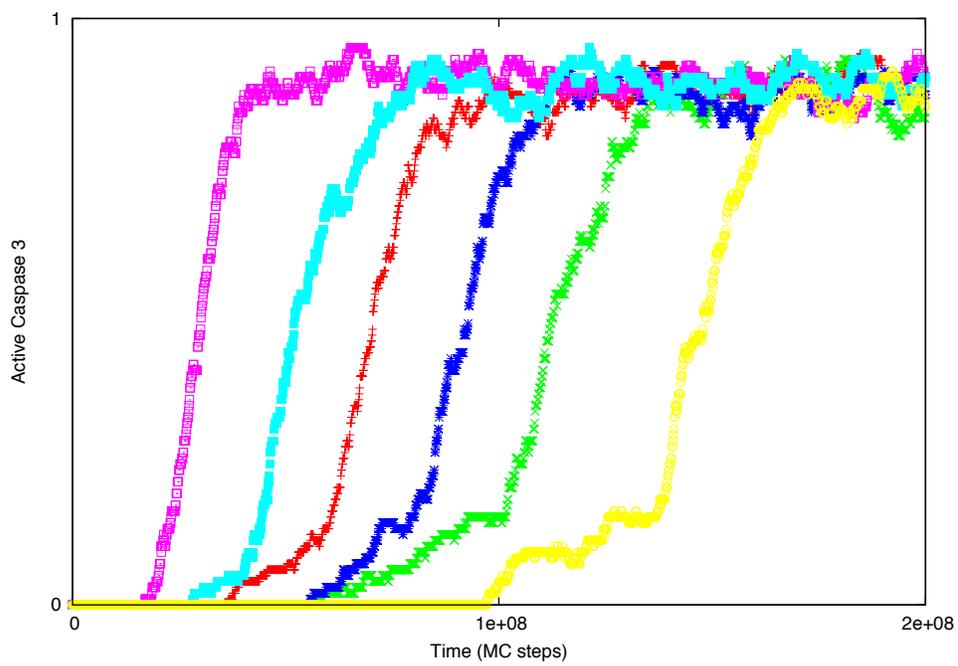

Figure 7a.

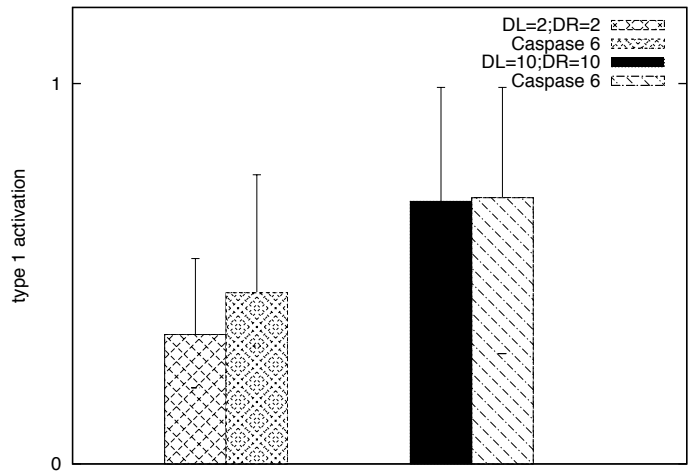

Figure 7b.

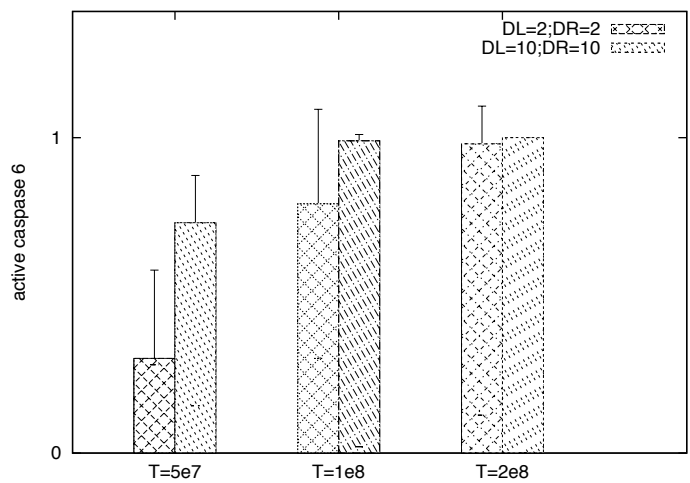

Figure 7c.

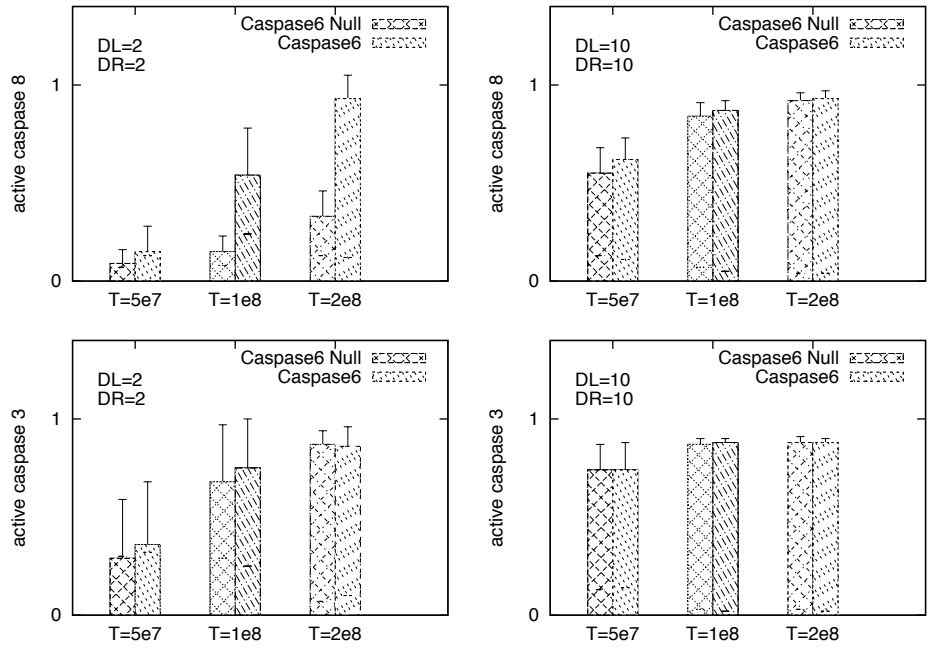

**Supplemental Figure 1.** Clustering of death receptors as the energy parameter (for death receptor clustering) is varied: (a) $E_{dd} = -2\ K_BT$ and (b) $E_{dd} = -3\ K_BT$. In each case, 2 representative single cell surfaces are shown for death ligand = 100 molecules (upper panel) and 2 molecules (lower panel). Death receptor concentration is kept fixed at 100 molecules.

(a)

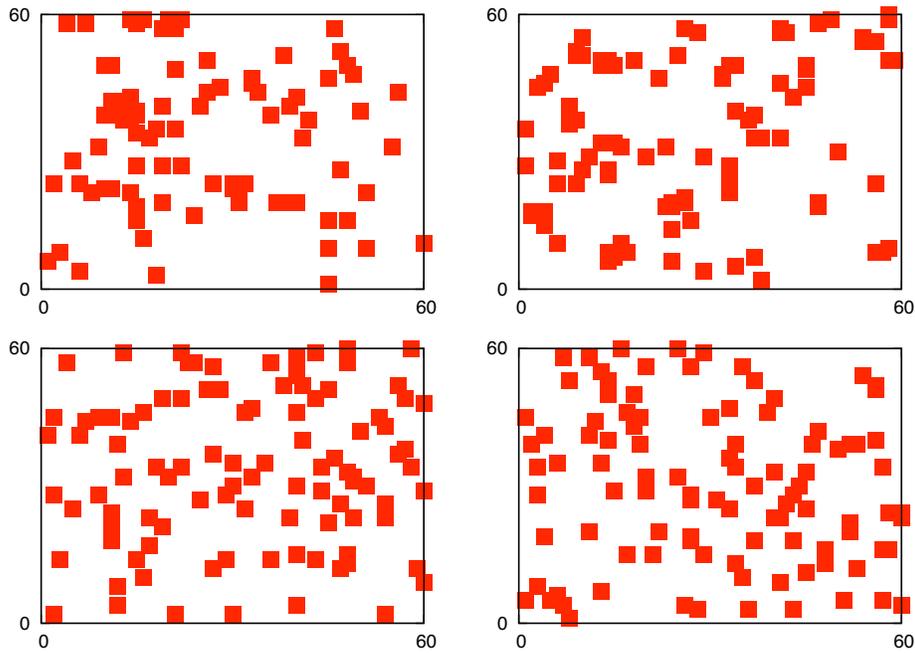

(b)

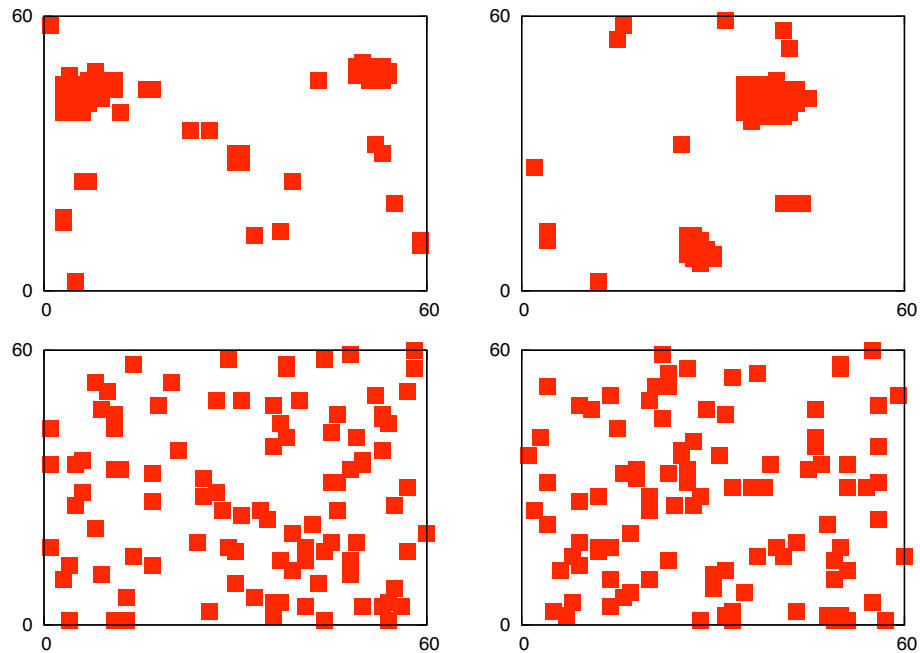